%&cp-aa
\def\ref{\par\noindent\hangindent 12pt}
\def\apj{ApJ,\ }
\def\aa{A\&A,\ }

\def\etal{{\rm et al.~}}
\def\Mpc{$h^{-1}${\rm Mpc}}
\def\ha{hereafter~}
\def\kms{km s$^{-1}$}
\font\small=cmr8
 1
\tolerance=4000
\vglue 0.2truein
% --> include POSTSCRIPT files:

\input epsf.tex
%\refereelayout           % a referee layout
\MAINTITLE={The Distribution of Galaxies in Voids}
\AUTHOR=
{ Ulrich Lindner@1, Maret Einasto@2, Jaan Einasto@2, 
          Wolfram Freudling@{3,4}, Klaus Fricke@1,
          Valentin Lipovetsky@{5}, Simon Pustilnik@{5}, 
          Yuri Izotov@{6}, Gotthard Richter@7}
\INSTITUTE 
{@1 Universit\"ats Sternwarte, Geismarlandstr. 11, D-37083 G\"ottingen,
Germany @2 Tartu Astrophysical Observatory, EE-2444 T\~oravere, Estonia
@3 Space Telescope -- European Coordinating Facility, D-85748 Garching, Germany
@4 European Southern Observatory, D-85748 Garching, Germany
@5 Special Astronomical Observatory,  Russian Academy of Sciences, 
357140 Nizhnij Arkhyz, Karachai-Cercessia, Russia
@6 Main Astronomical Observatory, Ukrainian Academy of Sciences,
Golossevo, Kiev, 252127, Ukraine
@7 Astrophysikalisches Institut Potsdam, D-14482 Potsdam, Germany }  

\ABSTRACT{We investigate the distribution of normal (faint) galaxies 
and blue compact galaxies (BCGs) in voids by analyzing their 
distribution as a function of distance from the void centers and by 
employing nearest neighbour statistics between objects of various 
subsamples. We find that galaxies in voids defined by brighter 
galaxies tend to be concentrated close to the walls of voids in a 
hierarchical manner, similar to the behavior of brighter galaxies.
The behavior of BCGs is in this respect similar to the one found for 
normal dwarf galaxies. The median nearest neighbour distance of BCGs 
from normal galaxies is approximately 0.7 \Mpc, which indicates that 
these galaxies are located in outlying parts of systems of galaxies 
defined by normal galaxies. }

 \KEYWORDS={ cosmology: observations 
--- large-scale structure of the Universe }   % must be given
\THESAURUS={12(12.03.3, 12.12.1)) }      % must be given
\OFFPRINTS={Ulrich Lindner, 
e-mail: ulindner \at uni-sw.gwdg.de}      % is optional
\DATE={Received 22 June 1995 / Accepted 22 March 1996} 
\maketitle
\MAINTITLERUNNINGHEAD{ Galaxies in Voids }
\AUTHORRUNNINGHEAD{ U. Lindner \etal }
\titlea {Introduction}
A remarkable property of the distribution of galaxies is the presence of
huge regions devoid of galaxies.  Examples of such voids are the Perseus
void behind the Perseus-Pisces supercluster (Einasto, J\~oeveer and Saar
1980), the Bootes void (Kirshner \etal 1981), and the Northern Local Void
between the Local, Coma, and Hercules superclusters (Freudling \etal 1988,
Lindner \etal 1995, \ha paper~I). 

In paper~I we have investigated void properties by considering regions 
completely empty of galaxies drawn from different luminosity limited 
samples up to $M=-18.8$ in the region of the Northern Local void. For 
different absolute magnitude limits, the empty regions were determined 
separately. Therefore, every luminosity limited sample defines a 
different set of ``voids''. We have argued that differences in the 
sizes of voids defined by varying luminosity limits are not only due 
to the differences in the number of galaxies, but describe a property 
of the spatial distribution of galaxies: voids defined by bright
galaxies contain galaxies which define smaller voids. These voids in
turn contain fainter galaxies which define yet smaller voids and so on.
In this sense both galaxies and voids form hierarchical systems. It is
not clear whether the hierarchy of the distribution of galaxies and
voids continues towards galaxies fainter than the luminosity limit
$M=-18.8$ used in paper~I.

A number of dwarf galaxies were found in  voids (Balzano \& Weedman
1982, Moody \etal 1987, Freudling \etal 1991, 1992, Peimbert \&
Torres-Peimbert 1992, Szomoru \etal 1993, Szomoru 1994 and references
therein). These galaxies are not distributed
homogeneously inside these voids, but they were found in the
vicinity of other galaxies inside voids (Szomoru 1994) or they are
concentrated to peripheral regions of voids  (Hopp 1994). These data
suggest that the hierarchy of the distribution of galaxies may continue
to fainter galaxies.
In most of these cases galaxies were studied in well-known distant
voids such as the Bootes void. However, it is difficult to detect 
normal dwarf galaxies at these distances and to study properties 
of the distribution of void galaxies in detail.

It has been suggested that voids are populated by galaxies which 
are in some respect peculiar (e.g. emission-line galaxies or blue 
compact dwarf galaxies). Pustilnik \etal (1994, 1995, hereafter PULTG) 
claimed that there exists a population of extremely isolated blue 
compact galaxies (\ha BCGs) in voids. This motivated us to investigate 
the distribution of the BCGs and compare it to the distribution of 
other galaxies. 

Our method is the following. First we investigate the distribution of 
galaxies inside voids which we have compiled in our previous 
work (paper~I) for the region of the Northern Local Void (\ha NLV) 
which is the closest cluster--defined void (i.e. supervoid) in our 
vicinity. In particular, we will investigate a possible continuation of
the void hierarchy to fainter galaxies. Next, we will study whether BCGs 
are exceptional in their distribution or if they follow the same 
distribution as normal galaxies of similar morphological type
and luminosity. For this purpose we investigate the distribution 
of BCGs in nearby voids using our void catalogues as well as the 
distribution of BCGs with respect to other (normal) galaxies using 
nearest neighbour statistics. In particular, we will try to clarify the 
question whether BCGs do form a particular population of relatively 
isolated galaxies in voids.

The paper is organized as follows. In \S 2 we describe the observational 
data used. In \S 3 we study the radial distribution of normal galaxies 
and BCGs inside of voids using void catalogues in the Northern Local 
Void region. In \S 4 we compare the distribution of normal galaxies of 
different absolute magnitude and BCGs. In \S 5 we use the nearest 
neighbour distribution test to compare the distribution of BCGs and 
normal galaxies in the window covered by the Second Byurakan Survey.  
Our results are summarized in \S 6.

Throughout this paper we use a Hubble constant of $H = h\cdot 100 \
{\rm km \ s}^{-1} \ {\rm Mpc}^{-1}$. All absolute magnitudes  are
calculated with $h=1$.

\titlea {The data} 
\titleb {Normal galaxies}
Void catalogues compiled in our previous work (paper~I) are 
available in the region of the Northern Local Void (\ha NLV). Our 
investigation on the distribution of BCGs is based on 
the Second Byurakan Survey (\ha SBS). We therefore constructed 
a sample of galaxies covering both, the NLV and SBS region. 

Galaxy redshifts were extracted from the compilation by Huchra (1994, 
ZCAT) and complemented with other published redshifts, mostly 21cm line 
redshift surveys in the region of the Northern Local Void (Freudling 
1995 and references therein). All sources of redshifts were carefully 
cross correlated to avoid duplicate entries from several sources in our 
final sample.
Distances were derived from the observed redshifts by correcting 
them for the solar motion, the Virgocentric flow, relativistic 
effects, and the velocity dispersion in clusters of galaxies. 
Absolute magnitudes were corrected for galactic extinction
as in Einasto \etal (1984). These galaxy samples are complete to 
$m = 14.5$ and fairly complete to $m = 15.5$ (see paper~I for details).  
 
The NLV is bounded by $12^h \leq \alpha \leq 18^h$ and $\delta \geq 0^\circ$
and our respective sample is 18000 \kms\ deep in redshift and comprises 
about 5500 objects. The galaxy subsample which overlaps with the BCGs 
from the Second Byurakan Survey is \ha referred to as the SBS window. 
This region is slightly bigger than the original area of our BCG sample 
in order to avoid edge effects. It is bounded by $6^h 30^m \leq \alpha 
\leq 18^h 30^m$ and $40^\circ \leq \delta \leq 70^\circ$. The galaxy 
sample in this SBS window is 12000 \kms\ deep and contains approximately 
2500 galaxies.

\titleb{Blue Compact Galaxies}
The sample of BCGs was primarily assembled from objective prism survey 
plates obtained with the 1-m Schmidt telescope at the Byurakan Observatory 
of the Armenian Academy of Sciences during the Second Byurakan Survey 
which covers the region $7^h 40^m \leq \alpha \leq 17^h 20^m$ and 
$49^\circ \leq \delta \leq 61.2^\circ$ on the sky. All objects showing 
strong or moderate emission lines were observed spectroscopically with 
the 6m telescope of the Special Astrophysical Observatory. The BCG sample
was constituted by selecting all emission--line galaxies with a H{\small II} 
region--like spectrum showing a narrow [O{\small III}] $\lambda 5007$ 
\AA\ line. For a discussion of further details we refer to PULTG.

This BCG sample was recently updated by one of us (VL)
and contains 210 objects in the redshift range $0 - 12000$ \kms.
Apparent magnitudes range from 12.5 to 19.5 with a distinct 
maximum near 18 and a steep cut--off for fainter magnitudes 
indicating the completeness limit. 
Absolute magnitudes range from $-11.5$ to $-21$, most
of them are situated between $-15$ and $-18$, which 
are typical luminosities of dwarf galaxies.

\titlea {Radial distribution of galaxies inside voids}
\titleb {Galaxy subsamples and void catalogues }
The method used in this section is to consider {\it regions completely 
empty of a certain type of objects}, in the following referred to as 
voids (cf. paper~I). This void definition is used throughout this paper. 
In this study we discuss only voids determined by galaxies of a given 
absolute magnitude limited sample. In this sense we say that respective 
voids are `surrounded' or `outlined' or `defined' by galaxies from the
respective sample.

In this section we investigate the distribution of galaxies inside voids 
from our void catalogues published in paper~I. To generate the void 
catalogues the sample of galaxies in the NLV region was divided into 
three volume-limited, cubic subsamples with side lengths $D_{lim}$ of 
60, 90 and 120 \Mpc, respectively (cf. paper~I). Number counts and 
absolute magnitude limit of each subsample, denoted NLV-M6, NLV-M9, 
and NLV-M12 respectively, are given in Table~1.

In order to study the distribution of galaxies inside voids we use 
samples with all known faint galaxies included instead of absolute 
magnitude limited samples. These samples are denoted as NLV--An 
(n=6,9 and 12) and are also listed in Table~1.

\begtabfull \tabcap{1}
{Subsamples selected from the sample of normal galaxies in 
the NLV region ($12^h \leq \alpha \leq 18^h$ 
and $\delta \geq 0^\circ$).
For further explanations see text }
\halign { # \hfil & \hfil # & \hfil # & \hfil # & \hfil #\cr
\noalign {\smallskip}
\noalign{\hrule}
\noalign{\smallskip}
Sample & $D_{lim}$ \ \ & \qquad $M_{lim}$ & \qquad $N$ \cr
name   & \qquad [\Mpc]  & &\cr
\noalign{\smallskip}
\noalign{\hrule}
\noalign {\medskip}
 NLV--M6  &   60 \qquad &  --18.8  &  1508 \cr
 NLV--M9  &   90 \qquad &  --19.7  &  1858 \cr
 NLV--M12 &  120 \qquad &  --20.3  &   877 \cr
\noalign{\smallskip}
 NLV--A6  &   60 \qquad &    --- \ &  2364 \cr
 NLV--A9  &   90 \qquad &    --- \ &  4049 \cr
 NLV--A12 &  120 \qquad &    --- \ &  4457 \cr
\noalign{\medskip}
\noalign{\hrule} } 
\endtab

Our void catalogues contain voids defined by galaxies of three 
absolute magnitude limited samples of different depth. Brightest 
galaxies (sample NLV--M12) define 37 voids, which have been compiled 
in catalogue A. Catalogue B contains 41 voids defined by galaxies 
from sample NLV--M9 and 25 voids defined by galaxies from sample 
NLV--M6 are listed in catalogue C.

By definition, these voids contain only galaxies fainter than the 
limiting magnitude used in the void definition. In the following we 
use the term {\it faint galaxies} for galaxies which are fainter than 
this limiting magnitude. A--, B--, and C--voids have different limiting 
absolute magnitudes used in their definition, thus the term ``faint 
galaxy'' has a {\it relative} meaning. Faint galaxies from samples 
NLV--An minus NLV--Mn (n = 6, 9 or 12) are possible candidates for 
galaxies in voids but most of them do not lie in voids because they 
are associated to the structures delineating the voids.

Mean diameters of voids listed in the A--, B--, and C--catalogues
are $22 \pm 1$, $16 \pm 1$ and $13 \pm 1$ \Mpc, respectively (paper~I, 
Table~4b). The decrease of void diameters is not only due to the different 
number of galaxies in the samples. There are two other relevant factors 
contributing to this trend. First, C--voids are located closer to the 
Virgo Supercluster, and voids in high--density regions are smaller than 
in low--density regions. Second, fainter galaxies surround systems of
bright galaxies and thus enter into voids defined by these bright 
galaxies (for details cf. paper~I).

\begfigwid 11.0cm
\vskip -16.5cm
\epsfysize=14.0cm
% \hskip -1.9cm
{\epsffile{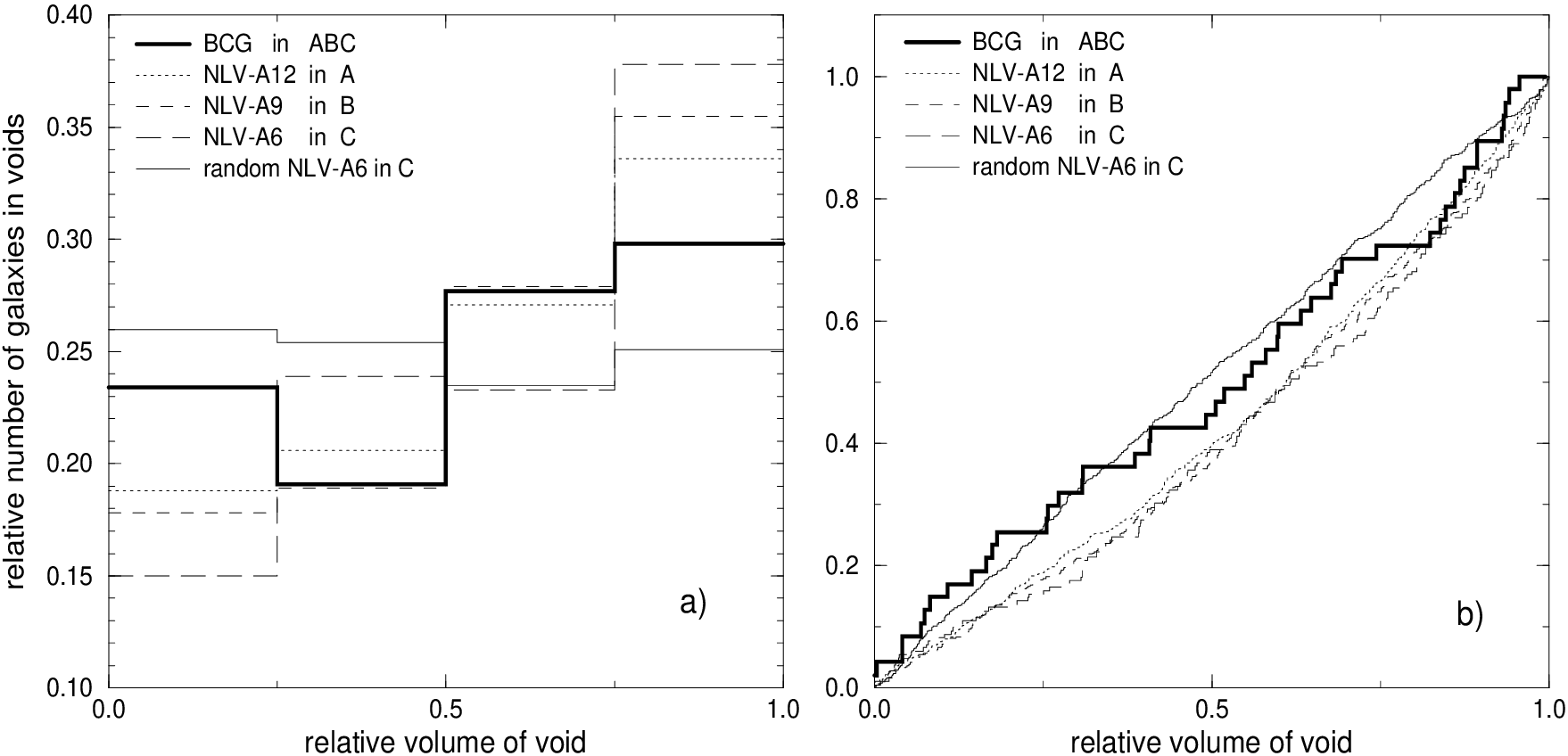}}
% \vskip -1.5cm 
\figure{1}
{Radial distribution of galaxies (samples NLV--An, see Table~1)
in voids from the A--, B-- and C--catalogues: {\bf a)} differential 
distribution binned in shells of equal volume, {\bf b)} cumulative
unbinned distribution. The thick solid line indicates the radial 
distribution of BCGs averaged over all three void catalogues. The 
radial distribution of random points according to the galaxy 
sample NLV--A6 in C--voids is also shown}
\endfig

\titleb{Normal galaxies in voids}
To study the radial distribution of galaxies in voids we use
samples of all galaxies in the NLV region of the sky (samples NLV-An)
which contain bright and faint galaxies as well. If a galaxy is located 
in a void, we express its distance from the void center in units of the
relative volume $(r/R)^3$, where $r$ is the distance of a particular
galaxy from the void center, and $R$ is the void radius. The
distributions of these distances from the void center were 
averaged for each void catalogue separately.

The results are presented in Figure~1. Figure~1a gives the differential
distribution of galaxies binned in relative internal void volume, and
Figure~1b gives the respective integral (cumulative and unbinned)
distribution. The distribution of galaxies from the samples NLV--A12, 
NLV--A9 and NLV--A6 in A--, B--, and C--voids is shown with dotted, 
dashed and long dashed lines, respectively. 

It can be seen that the relative radial distribution of galaxies 
in voids is practically identical for all three void catalogues. 
We stress that the properties of voids from these three catalogues 
are quite different: voids are defined by galaxies from three different 
absolute magnitude limits, they cover different volume in depth and 
have different size. The properties of the samples of galaxies inside 
voids from  these three classes are also very different. They have 
different upper absolute magnitude limits and therefore different 
level of incompleteness because in nearby samples we have more 
fainter galaxies than in faraway ones.

A Kolmogorov--Smirnov (\ha KS) test shows that the probability that all
three distributions are taken from the same parent sample, lies between
0.52 and 0.89 (see Table~2). Therefore, we conclude that on average the
relative radial distribution of galaxies in voids is almost independent
of the environment and the mean size of the voids and the galaxy
properties as well.  In paper~I only galaxies brighter than $-18.8$
were used to establish the hierarchical picture of the distribution of
galaxies and voids. Now this similarity suggests that this hierarchy is
not limited to the absolute magnitude $-18.8$, but continues towards 
fainter galaxies.

The shape of curves indicates that most faint galaxies in voids are 
located in the outer region, i.e. close to systems of galaxies which 
form void boundaries, and a smaller fraction (about one third) is 
situated in central regions of voids. Similar results were 
obtained by Hopp (1994).

\begtabfull \tabcap{2} 
{Results of the Kolmogorov--Smirnov test to compare the radial
distribution of BCGs and (normal) galaxies in voids as
shown in Figure~1b).
$\beta$ is the probability that distributions compared are taken from
the same parent distribution. }
\halign { # \hfil & # \hfil  & \hfil # \hfil  \cr
\noalign {\smallskip}
\noalign{\hrule}
\noalign{\smallskip}
1st distribution~~~~~& 2nd distribution~~~~~ & $\beta$ \cr
\noalign{\smallskip}
\noalign{\hrule}
\noalign {\medskip}
 NLV--A12 in A  &  NLV--A9 in B     & 0.875 \cr
 NLV--A12 in A  &  NLV--A6 in C     & 0.523 \cr
 NLV--A9 in B   &  NLV--A6 in C     & 0.886 \cr
\noalign {\medskip}
 BCG in ABC & NLV--A12 in A & 0.685 \cr
 BCG in ABC & NLV--A9 in B  & 0.464 \cr
 BCG in ABC & NLV--A6 in C  & 0.386 \cr
 BCG in ABC & random NLV--A6 in C \quad & 0.465 \cr
% BCG in ABC & PRD6$_C$& 0.649 \cr
\noalign {\medskip}
 NLV--A6 in C  & random NLV--A6 in C      & 0.000 \cr
% NLV--A6 in C  & PRD6$_C$     & 0.000 \cr
% PNLV--A6 in C & PRD6$_C$     & 0.882 \cr
\noalign{\medskip}
\noalign{\hrule} } 
\endtab

For comparison we generated catalogues of random points with the same
number of objects and the same distance distribution of the number
density of objects as for the sample NLV--A6. The dependence of the 
number density on distance is mainly due to selection effects, i.e. 
only nearby intrinsically faint galaxies can be detected. In addition, 
the decreasing density with increasing distance from the Virgo 
Supercluster contributes to the radial distribution of galaxies.
We consider only distances up to $60$ \Mpc\ and consequently galaxies 
in C--voids. The random sample is labeled as ``random NLV--A6 in C''
in Figure~1 and Table~2.

\begfigwid 14.0cm
\vskip -14.0cm
\epsfysize=14.0cm
\hskip -1.9cm
{\epsffile{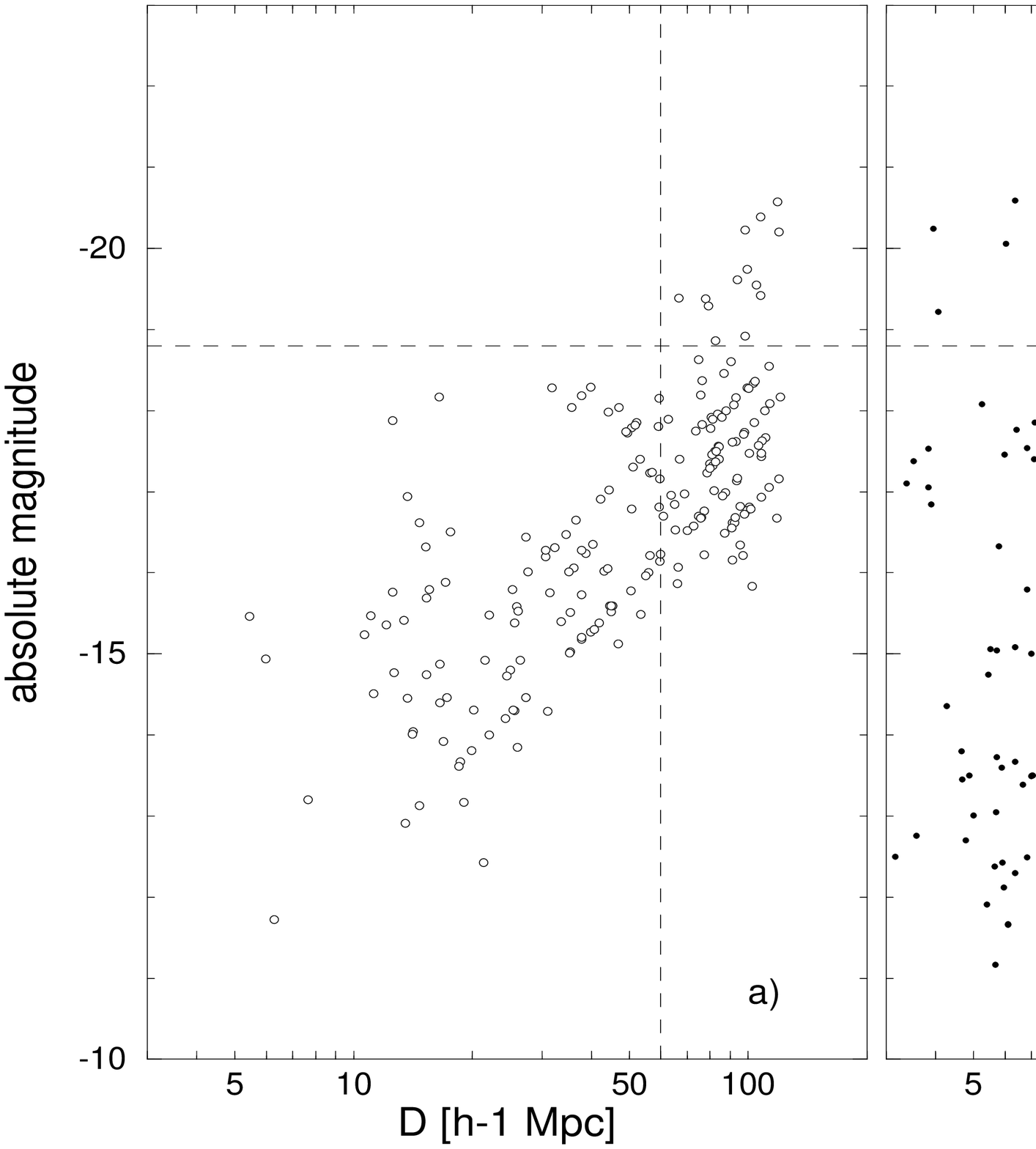}}
\vskip -1.5cm 
\figure{2}
{Absolute magnitude as a function of distance $D$ (in \Mpc) 
for {\bf a)} BCGs and {\bf b)} 
all normal galaxies (sample RA12). Horizontal and vertical
dashed lines indicate the appropriate absolute magnitude limit
$M_{lim} = -18.8$ and distance limit $D_{lim} = 60$ \Mpc} 
\endfig

The distribution of random points differs significantly from the
distribution of real galaxies; the KS-probability that distributions
are taken from the same parent sample is smaller than $10^{-3}$ (cf.
Table~2).

\medskip\noindent
From Figure~1 and the above discussion we can draw the following
three conclusions. 

\item{$\bullet$}
Voids contain galaxies fainter than the absolute magnitude 
limit used in the void definition.

\item{$\bullet$}
Galaxies inside voids are far from being randomly distributed.  Most of
them are concentrated to the rims of the voids, but about one third of
galaxies are also found close to the central regions of  voids.

\item{$\bullet$}
The relative distribution of galaxies in voids defined by galaxies of
different absolute magnitude is practically identical. 

\noindent
The last conclusion is somewhat surprising since the three void catalogues 
are different in depth and mean size of voids, and the results are based 
on samples of different incompleteness. Since it is unlikely that the 
differences in the incompleteness conspire to transform different radial 
distributions into similar curves in Figure~1, we conclude that the 
intrinsic radial distribution of the considered galaxies is similar and 
that incompleteness does not significantly affect this measurement. It 
therefore appears that the faint galaxies in voids defined by brighter 
galaxies are distributed similarly for different magnitude limits. This 
strongly suggests that the hierarchical distribution of galaxies and 
voids found in paper~I down to an absolute magnitude limit $-18.8$ 
continues to fainter luminosity limits.

\titleb {BCGs in voids}
The NLV region covers only half of the SBS window. Therefore
the number of BCGs in voids from our catalogues is very small: there
are 16 objects in voids from the A--catalogue, 25 BCGs reside in
B--voids, and only 6 BCGs in C--voids.  Thus we have averaged over 
all three void catalogues to determine the radial distribution of 
BCGs in voids, indicated by bold solid lines in Figure~1.

Figure~1b shows that the cumulative radial distribution of BCGs 
lies between the distribution of random points and normal galaxies.
This qualitative impression is confirmed by the KS test: the
probability that BCGs are distributed like normal galaxies lies 
between 0.39 and 0.68, and that BCGs are distributed as random 
points is $0.46$ (see Table~2).

There is no evidence that the distribution of BCGs differs from
a random distribution, but the statistics is poor. Therefore we
shall perform further tests which are potentially more sensitive to
detect deviations from a random distribution.

\begfigwid 22cm 
\vskip -25.7cm
\hskip -1.0cm
\epsfysize=30.0cm
{\epsffile{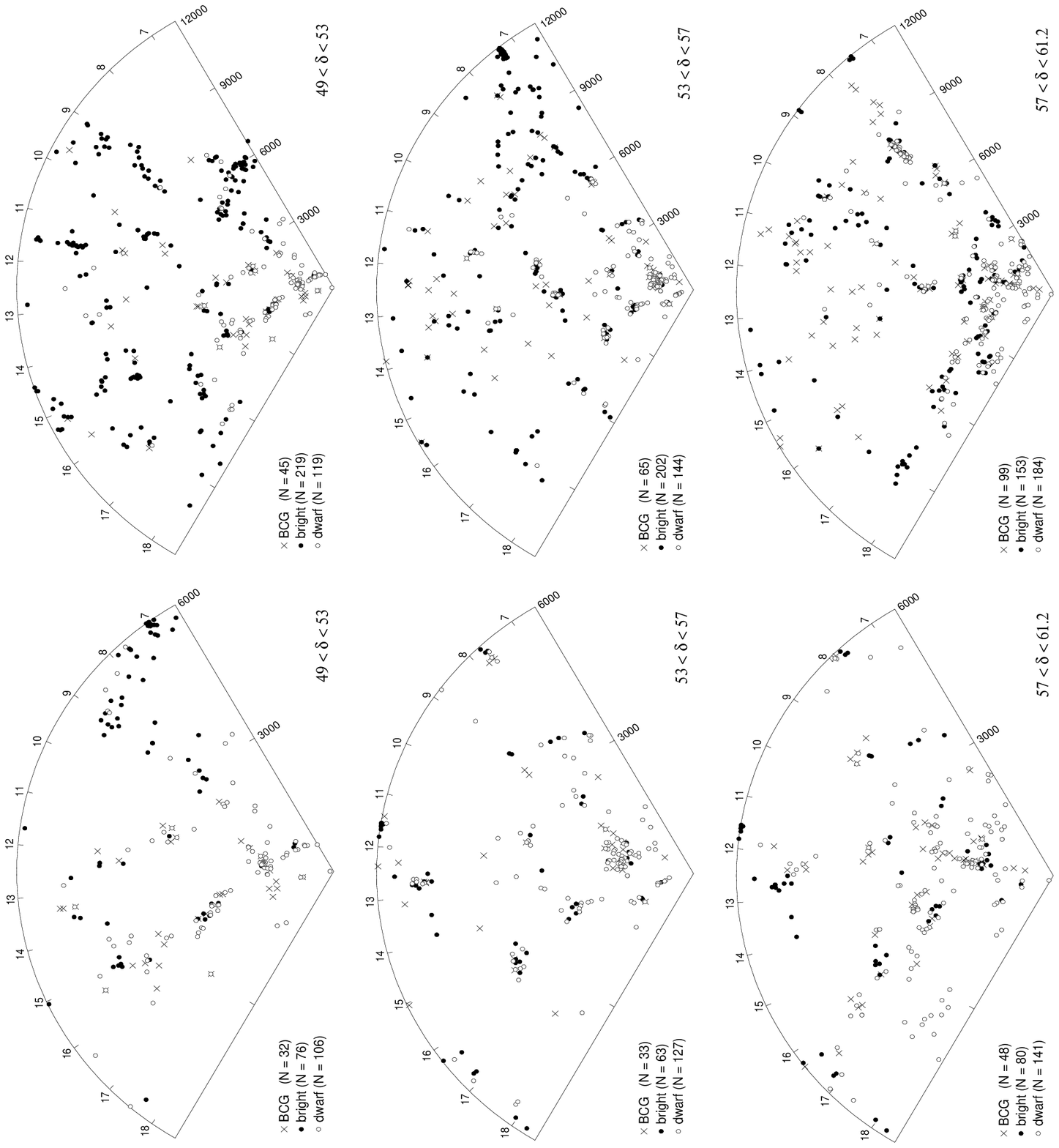}}
\vskip -7.5cm
\figure{3}
{Wedge diagrams of the SBS window on sky ($7^h 40^m \leq \alpha 
\leq 17^h 20^m$, $49^\circ \leq \delta \leq 61.2^\circ$) showing 
the distribution of BCGs, normal dwarf galaxies ($M > -18.8$) 
and bright galaxies ($M \leq -18.8$) in polar equatorial 
coordinates up to a depth of 60~\Mpc\ and 120~\Mpc, respectively}
\endfig

\titlea{The distribution of BCGs and normal galaxies}
Our study of the radial distribution of normal galaxies and BCGs in 
voids so far ignores properties of the mutual distribution of galaxies 
relative to each other. Now we focus our attention to the comparison 
of the distribution of BCGs with respect to normal galaxies. First we 
give a cosmographic description and subsequently in \S 5 we shall 
perform a quantitative study using nearest neighbour statistics.

\smallskip
\titleb {The distribution of distances and absolute magnitudes}
The distribution of BCGs and normal galaxies of different absolute
magnitude in the SBS window at various distances is shown in
Figure~2.  This distribution shows that the absolute magnitude range of
BCGs and normal galaxies depends on the distance, as expected
for apparent magnitude limited samples.

Up to a distance of 60 \Mpc\ there are no BCGs brighter than $-18.8$.
On the other hand, the number of normal galaxies in our sample at
distance larger than 60 \Mpc\ and fainter than absolute magnitude
$-18.8$ is very small. Thus, we can use the distance limit 60
\Mpc\ as a suitable limit to compare the distributions of BCGs and
normal galaxies. Also we can consider normal galaxies of luminosities
not exceeding this absolute magnitude as suitable comparison galaxies
for BCGs.

For the following we use the term {\it dwarf galaxies} to designate
galaxies fainter than $M = -18.8$. The absolute magnitude limit
$-18.8$ corresponds to an apparent magnitude 15.5 at the limiting
distance of 60 \Mpc. Respective absolute magnitude and distance limits
are shown in Figure~2 by dashed lines. The limit $M = -18.8$
corresponds also to the faintest limit of absolute magnitudes of
galaxies used in the construction of the void catalogues in paper~I,
which is useful to compare results obtained in different sections of
this study.

\titleb {Wedge diagrams of galaxies}
In Figure~3 we show wedge diagrams for various declination ranges in
the SBS window, illustrating the distribution of BCGs and normal dwarf
and normal galaxies in equatorial coordinates. Two sets of diagrams are
presented, with maximum distance 60~\Mpc\ and 120~\Mpc, respectively.
These plots show the filamentary distribution of galaxies and the
presence of voids. We see that BCGs are not very isolated. Most of 
them are forming continuations of filaments of normal galaxies or 
they constitute bridges between systems of galaxies. The visual
impression from these figures is that the distribution of BCGs
is not very different from the distribution of other galaxies of
similar luminosity.

There are almost no dwarf galaxies near BCGs at distances exceeding 
60~\Mpc. Even in the nearby region ($D \leq 60$ \Mpc) the fraction 
of the number of dwarf galaxies to the number of all galaxies decreases 
rapidly with distance. This is a clear manifestation of selection 
effects, already discussed along with Figure~2.

\begtabfull \tabcap{3}
{Subsamples selected from the galaxy sample in 
the SBS window ($6^h 30^m \leq \alpha \leq 18^h 30^m$ 
and $40^\circ \leq \delta \leq 70^\circ$)
used in the nearest neighbour analysis.
For further explanations see text }
\halign { # \hfil & \hfil # & \hfil # & \hfil # & \hfil #\cr
\noalign {\smallskip}
\noalign{\hrule}
\noalign{\smallskip}
Sample & $D_{lim}$ \ \ & \qquad $M_{lim}$ & \qquad $N$ \cr
name   & [\Mpc]        &                  &            \cr
\noalign{\smallskip}
\noalign{\hrule}
\noalign {\medskip}
 BCG3  &   30 \qquad & --- \  & \ 54 \cr
 BCG4  &   40 \qquad & --- \  & \ 77 \cr
 BCG6  &   60 \qquad & --- \  &  113 \cr
 BCG8  &   80 \qquad & --- \  &  138 \cr
 BCG10 &  100 \qquad & --- \  &  184 \cr
 BCG12 &  120 \qquad & --- \  &  209 \cr
\noalign{\smallskip}
 RM3   &   30 \qquad & --17.3  &  325 \cr
 RM4   &   40 \qquad & --17.9  &  372 \cr
 RM6   &   60 \qquad & --18.8  &  450 \cr
 RM8   &   80 \qquad & --19.4  &  474 \cr
 RM10  &  100 \qquad & --19.9  &  466 \cr
 RM12  &  120 \qquad & --20.3  &  309 \cr
\noalign{\smallskip}
 RA3   &   30 \qquad & --- \  &         685 \cr
 RA4   &   40 \qquad & --- \  &         909 \cr
 RA6   &   60 \qquad & --- \  &        1301 \cr
 RA8   &   80 \qquad & --- \  &        1727 \cr
 RA10  &  100 \qquad & --- \  &        2204 \cr
 RA12  &  120 \qquad & --- \  & \qquad 2493 \cr
\noalign{\smallskip}
 RD3  &   30 \qquad & --18.8  &  556 \cr
 RD4  &   40 \qquad & --18.8  &  696 \cr
 RD6  &   60 \qquad & --18.8  &  871 \cr
 RD8  &   80 \qquad & --18.8  &  960 \cr
 RD10 &  100 \qquad & --18.8  & 1000 \cr
 RD12 &  120 \qquad & --18.8  & 1008 \cr
\noalign{\smallskip}
 RG3  &   30 \qquad & --20.3  &  22 \cr
 RG4  &   40 \qquad & --20.3  &  30 \cr
 RG6  &   60 \qquad & --20.3  &  50 \cr
 RG8  &   80 \qquad & --20.3  & 113 \cr
 RG10 &  100 \qquad & --20.3  & 221 \cr
 RG12 &  120 \qquad & --20.3  & 309 \cr
\noalign{\medskip}
\noalign{\hrule} } 
\endtab

\begfigwid 16.0cm
\vskip -18.5cm
\epsfysize=16cm
\hskip -0.8cm
{\epsffile{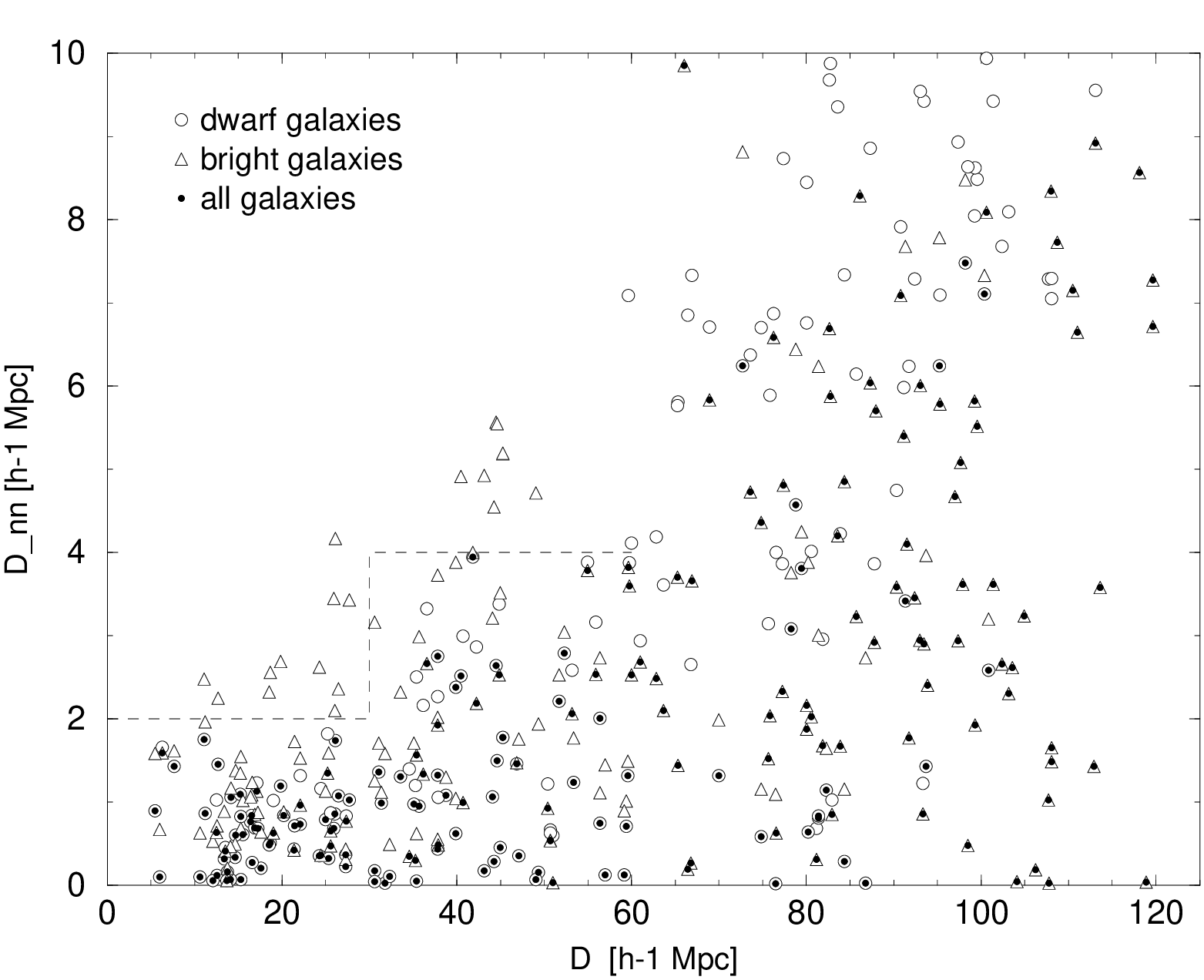}}
\vskip -0.1cm 
\figure{4}
{Nearest neighbour distances ($D_{nn}$) for individual BCGs
versus normal galaxies
as a function of distance, $D$, from the observer. 
Dots indicate NN distances searched among {\it all galaxies} 
(sample RA12), open circles those searched 
only among (normal) {\it dwarf galaxies} (sample RD12) 
and open triangles indicate NN distances searched among
the remaining {\it bright galaxies} (sample RA12 - RD12)}
\endfig

\titlea {The nearest neighbour test}
The cosmographic approach in the last section indicates that the 
distribution of BCGs is not very different from the distribution 
of other galaxies of similar luminosity. To get a more accurate
picture we apply in this section the nearest neighbour test. 

For each galaxy of a {\it test sample} a nearest neighbour (\ha NN) 
is searched among galaxies from a {\it reference sample}. As test 
samples BCGs and special selections of normal galaxies were used. 
Various subsamples of normal galaxies were used as references.
BCGs as well as normal galaxies were taken from the SBS window.

\bigskip
\titleb {Subsamples of BCGs and normal galaxies}
For the purpose of this study, we divided the samples of BCGs and 
normal galaxies into subsamples of different absolute magnitude limits.
A complete list of all subsamples used is given in Table~3. $D_{lim}$ 
denotes the sample depth and $N$ gives the number of galaxies in
respective subsamples.

Samples of blue compact galaxies are designated as BCGn, where n is
the limiting distance in tens of Megaparsecs. No absolute magnitude 
limit $M_{lim}$ is given, i.e. our BCG samples are not complete.

For the comparison of the distribution of BCGs with normal galaxies 
four different kinds of subsamples in the SBS window are used. 
Contrary to the (cubic) subsamples from the NLV region these samples 
are conical in shape. The four groups of subsamples are named RMn, 
RAn, RDn and RGn (R for {\it reference sample}), where $n = 3, 4, 6, 8, 
10$ and $12$ denotes the sample depth ($D_{lim}$) in tens of Megaparsecs.
Absolute magnitude limited subsamples are designated as RMn. The
absolute magnitude limit $M_{lim}$ corresponds to the apparent
magnitude limit $15.5$ at the distance $D_{lim}$. Subsamples denoted 
as RAn are not absolute magnitude limited, i.e. they contain all normal 
galaxies with known redshifts from our compilation in the SBS window. 
These samples are not complete. The samples of dwarf galaxies RDn have 
a fixed upper luminosity limit $M_{lim} = -18.8$. RGn denotes subsamples 
of {\it giant} galaxies brighter than $M_{lim} = -20.3$.

\titleb {Nearest neighbours of BCGs}
In this section we aim to clarify which types of galaxies the BCGs 
are closest to. Presumably normal dwarf galaxies are most similar to 
BCGs. Therefore, the spatial distribution of the dwarf galaxies in the 
RA sample and the BCG sample should be similar. To test this hypothesis 
we find nearest neighbours of BCGs among three different galaxy samples,
the whole reference sample RA12, the subsample of dwarf galaxies,
RD12, and the complementary subsample of remaining bright galaxies,
RA12 -- RD12. The division between these two subsamples has the same
absolute magnitude value ($-18.8$) as the practical absolute magnitude
limit of BCGs within the 60 \Mpc\ sphere as discussed along with
Figure~2 in \S 4.1. 

\begfigwid 15.0cm
\vskip -18.5cm
\epsfysize=16.0cm
\hskip -0.2cm
{\epsffile{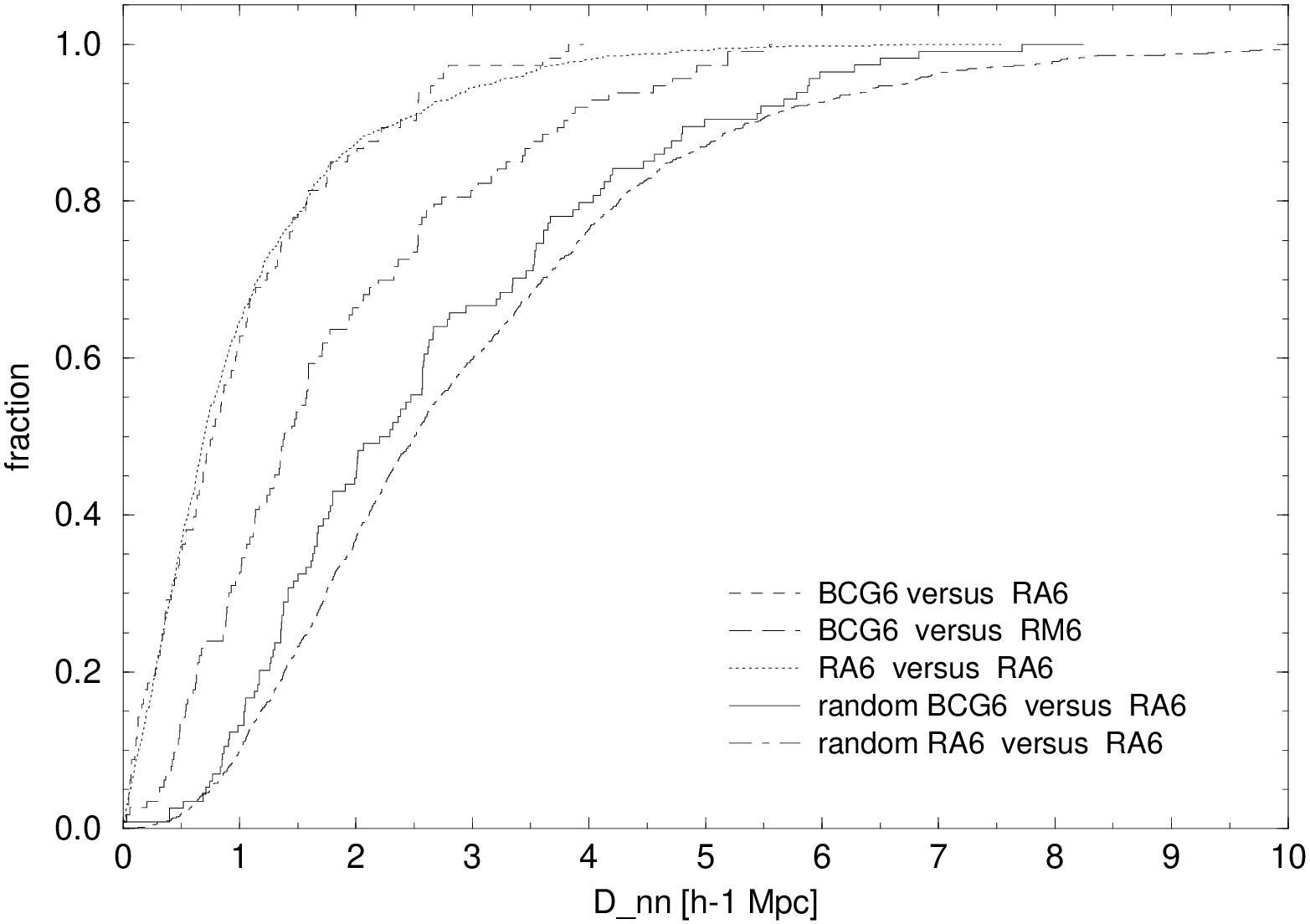}}
%\vskip -1.0cm 
\figure{5}
{Cumulative distributions of nearest neighbour distances $D_{nn}$ 
given in \Mpc\ for various real and random samples}
\endfig

In Figure~4 we give these NN distances, $D_{nn}$, as a function of the
distance, $D$, of BCGs from the observer. Distances of nearest
neighbours of BCGs were searched among all galaxies from the reference
sample (dots), or among the sample of dwarf galaxies or bright galaxies
(open circles and triangles, respectively). Coincidence of dots with
one of the other symbols indicates that the nearest neighbour was among
one of the specified type of galaxies. This plot gives a quick visual
impression of whether BCGs are located closer to dwarf or to bright
galaxies.

We see that for nearby BCGs, i.e. objects with distance less than 
60 \Mpc, most symbols for galaxies from the dwarf galaxy sample (RD12) 
and the whole galaxy sample (RA12) coincide. 
This indicates that most (71 \%) nearest neighbours of BCGs are 
dwarf galaxies. On larger distances most (80 \%) BCGs appear to 
have nearest neighbours preferentially among giant galaxies.
In particular Figure~4 shows that the distances of nearest neighbours 
are rather small (less than 2 \Mpc) up to a distance of about 30 \Mpc\ 
from the observer. In the distance interval from 30 to 60 \Mpc\ NN 
distances gradually increase up to 4 \Mpc. At larger distances from the 
observer NN distances are still higher and reach up to 10 \Mpc\ and more.

Such distance dependence is due to a selection effect as the fraction
of dwarf galaxies in our sample RA decreases with distance, as can be 
seen from Figure~2. 
Since our nearby sample is also incomplete for faint dwarf galaxies it 
may be possible that nearest neighbours of all BCGs are dwarf galaxies.

\titleb {The distribution of the NN distances of BCGs}
Now we consider the distributions of NN distances quantitatively.
Contrary to the discussion of the previous section, this investigation
is carried out on samples $60$ \Mpc\ deep. We determine NN
distances for test samples BCG6 and RA6. As reference samples we use
RA6 (all galaxies) and RM6 (absolute magnitude limited sample). We
compare these distributions also with the NN distributions of random
points. For this purpose random samples were generated with the same
number of objects and the same radial density distribution from the
observer as in the real BCG6 and RA6 samples.

The results of our calculations are plotted in Figure~5.  
Different line types denote NN distance distributions for 
various subsamples. The most striking message from the figure 
is that NN distance distributions of BCGs in respect to galaxies
from RA, RM and random samples are very different. The KS test
indicates, that these distributions are different at confidence 
level 99.9 \%. On the contrary, the NN distribution of BCGs in 
respect to RA6 galaxies is practically identical with the distribution 
of NN distances for galaxies from sample RA6 with themselves 
(KS probability is $\approx 0.76$).

Three conclusions may be drawn from this test. First, the distribution 
of NNs of all galaxies as well as of BCGs is very different from the 
distribution of randomly located points. In \S 3.3. we found that the 
{\it radial} distribution of BCGs in voids is almost the same as in 
random samples, i.e. the fraction of BCGs in central regions of voids 
is approximately proportional to the volume fraction. The NN analysis 
shows that the {\it relative} distribution of these galaxies is different.
All galaxies find close neighbours among other galaxies much closer than
randomly located points do. This result reflects the well-known fact 
that the galaxy distribution is clumpy or clustered, which is usually 
described by a correlation function.

Second, the tests show that the distribution of BCGs is 
practically indistinguishable from the distribution of 
normal galaxies of all luminosities (sample RA).

The third conclusion concerns the difference between nearest 
neighbours searched among galaxies from absolute magnitude 
limited samples and the respective total galaxy samples. 
This can best be studied quantitatively by median values of 
the NN distances for the samples of blue compact dwarf galaxies 
(BCGn) and normal dwarfs (RDn) in respect to the samples RMn
and RAn. Here we use also samples of depth up to  120 \Mpc. 
The results are presented in Figure~6.

\begfig 8.0cm
\vskip -9.3cm
\epsfysize=8.0cm
\hskip -0.8cm
{\epsffile{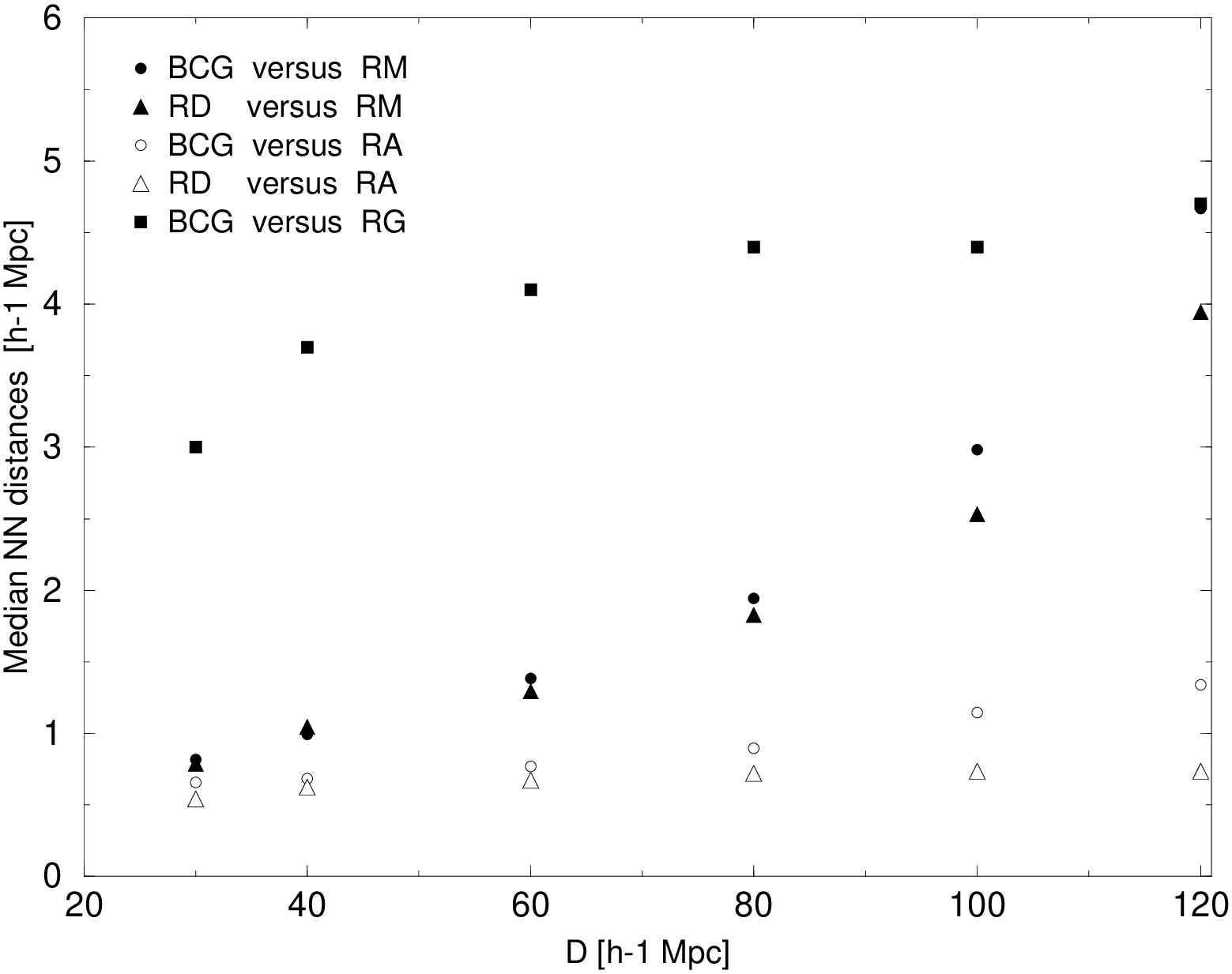}}
%\vskip -1.0cm 
\figure{6}
{Median NN distances of BCGs (circles) and normal dwarf galaxies 
(RD) (triangles) in respect to samples of all galaxies (RA) (open
symbols) and absolute magnitude limited galaxy samples (RM)
(filled symbols). 
Squares show median NN distances of BCGs from galaxies with the fixed
absolute magnitude limit $M = -20.3$ (sample RG). 
The depth $D$ of respective subsamples is used as an argument}
\endfig

\begfigwid 13.9cm
\vskip -16.9cm
\epsfysize=12.7cm
%\hskip 0.5cm
{\epsffile{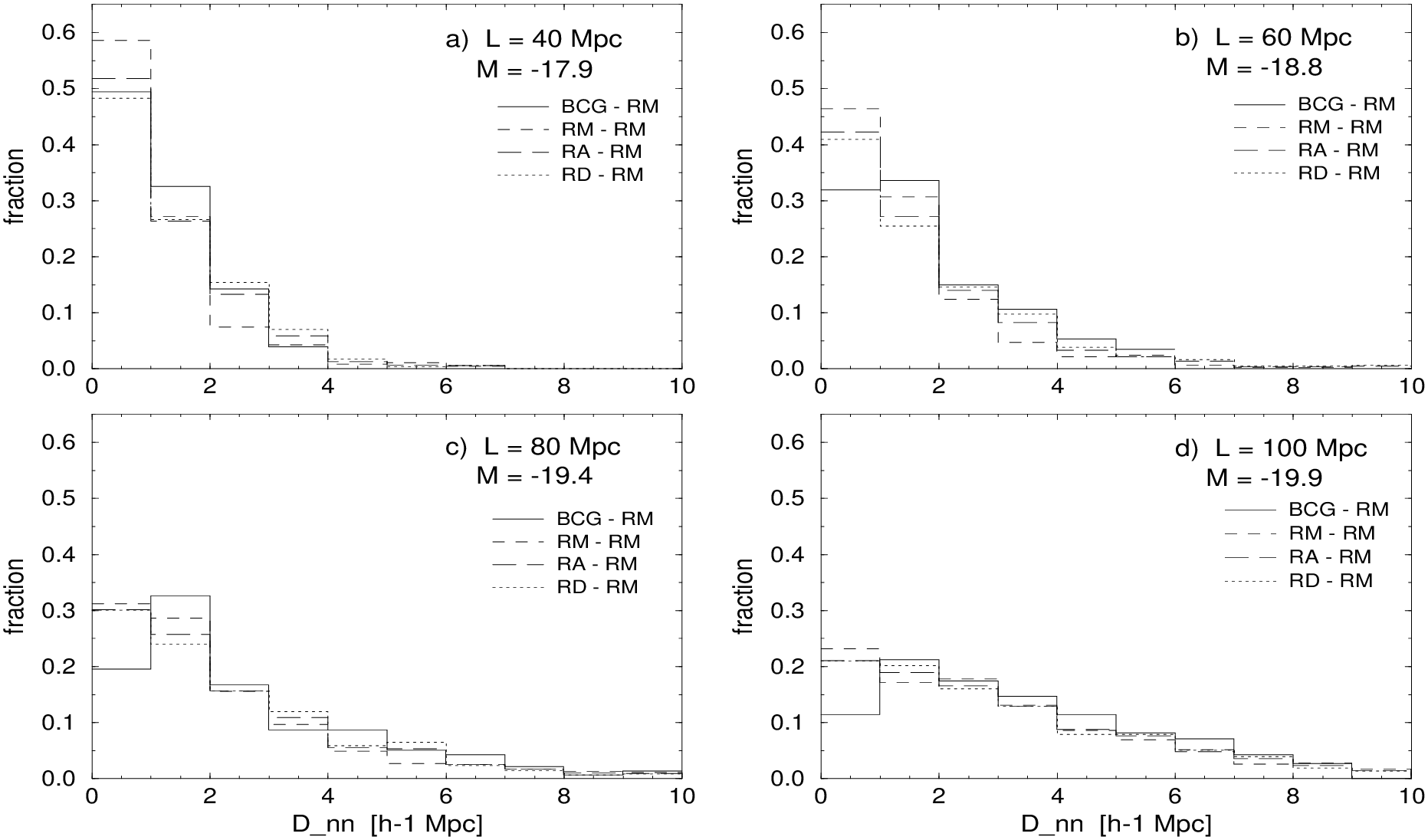}}
\vskip 0.3cm 
\figure{7}
{Differential distributions of nearest neighbour distances $D_{nn}$.
Each panel shows four histograms for test samples BCG, RM, RA and 
RD. RM is used as reference sample throughout. Panels a) through
d) are for subsamples of limiting distances 40, 60, 80, and 100 
\Mpc, with corresponding absolute magnitude limits a) $-17.9$, 
b)  $-18.8$, c) $-19.4$ and d) $-19.9$. Note that in panel c) and
d) curves for RA -- RM and RD -- RM coincide in the first bin }
\endfig
\begfigwid 14.3cm
\vskip -17.4cm
\epsfysize=12.6cm
\hskip -0.5cm
{\epsffile{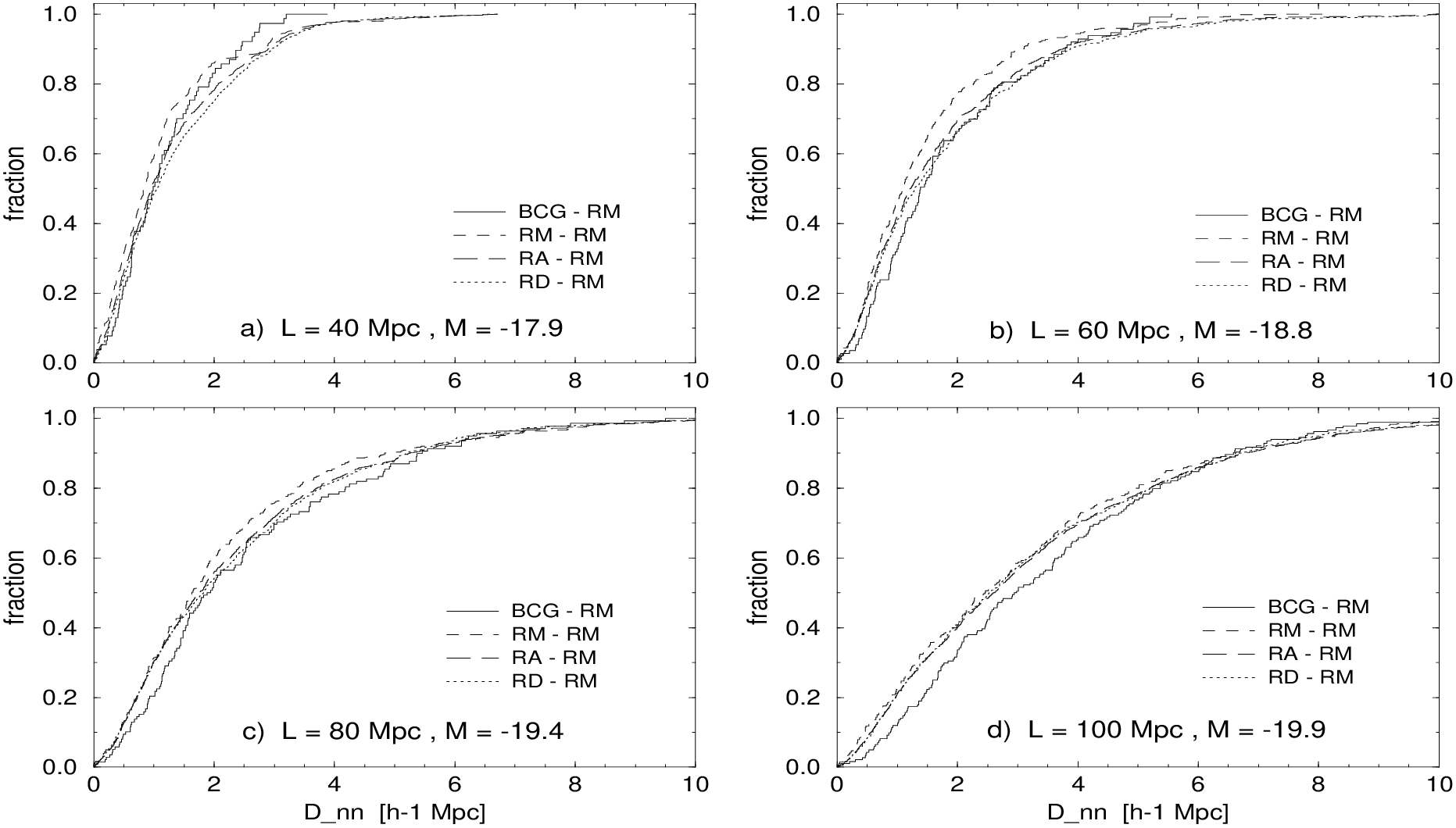}}
\vskip 0.3cm 
\figure{8}
{Cumulative distributions of nearest neighbour 
distances $D_{nn}$. All samples are the same as in Fig.7}
\endfig

The median distance of NNs of blue compact galaxies with respect to all
normal galaxies increases slowly with distance from the observer. This
small increase is due to a selection effect -- the number-density of
faint galaxies decreases with distance. The true median distance of NNs
of BCGs can be estimated from the nearby samples. The median distance
of NNs of BCGs determined for all galaxies from the sample RA6 (60
\Mpc\ deep) is 0.77 \Mpc\ and for the sample RA3 (30 \Mpc\ deep) 0.66
\Mpc. Since the nearby sample is also incomplete we must consider the
last number as an upper limit of the true median NN distance of BCGs
from normal (dwarf) galaxies.

The median distance of NNs searched between BCGs and galaxies from
bright absolute magnitude limited samples is 1.4 \Mpc\ and 0.8 \Mpc\
for samples RM6 and RM3, respectively. This shows that BCGs are located
from bright galaxies at larger distances than from dwarf ones.
We give also median NN distances of BCGs from very bright
galaxies to address the question of isolated galaxies in voids. 

PULTG have studied the NN distributions of BCGs in respect to 
normal bright galaxies from the CfA1. They claimed that 
$\approx 20$\% BCGs are situated in voids outlined by CfA1 
galaxies. To compare our results with those from PULTG we
selected a sample of very bright galaxies with absolute magnitude 
limit $-20.3$. This limit was chosen because it corresponds 
approximately to the apparent magnitude limit $14.5$ of the 
CfA1 survey at the distance where most if the isolated BCGs 
were found by PULTG. Subsamples from this sample of very bright 
reference galaxies are named RGn and are listed in Table~3.
Naturally RG12 coincides with RM12.   

NN distances from these very bright galaxies are considerably larger
than from less luminous galaxies, at all distance classes studied. Our
study shows that practically all BCGs have close neighbours among
dwarf galaxies if observational conditions (proximity to us) allow to
observe faint objects. We come to the conclusion that the result of
PULTG on the presence of a population of very isolated BCGs is due to
the use of a sample of very bright comparison galaxies.

The median distance of about 0.7 \Mpc\ is characteristic for galaxies
in groups and filaments. Most groups of galaxies have main galaxies
brighter than the absolute magnitude limit $-18.8$ considered here
(Vennik 1984, 1986) and the median distance from these galaxies is
larger, 0.9 \Mpc. In other words, our study emphasizes that BCGs are
located either in loose systems of galaxies defined by ordinary faint
galaxies, or in peripheral regions of groups of galaxies defined by
bright galaxies.

We conclude this section with a discussion of the physical meaning of
results on median NN distances presented in Figure~6. We try to answer
questions: Do results presented depend critically on the number of
particles in samples rather that on physical properties of these
samples? What is the difference between real samples and samples of
randomly distributed particles?

Median distances of BCGs were found in respect to the following
comparison samples: all galaxies (RA), very bright galaxies (RG), and
absolute magnitude limited samples of variable limit (RM). The number
of objects in RA samples increases with the increase of the sample
depth over 3 times, in the case of RG samples over 10 times, and is
approximately constant for RM samples (see Table~3). Respective median
NN distances are approximately constant in the first two cases, and
increase in the last case about 7 times. We see that there is no clear
dependence on the number of particles in the reference sample. On the
contrary, there is a very clear dependence on physical properties of
reference samples: physical properties (absolute magnitude interval
covered by samples) of samples RA and RG are essentially identical for
all distance classes, but different for the RM sample.  We see a direct
relationship between properties characterizing the distribution of objects
(median NN distances) and physical parameters of samples, and no clear
dependence  on the number of particles in respective samples.

\begtabfull \tabcap{4} 
{Results of the Kolmogorov--Smirnov test to compare the NN
distance distributions of BCGs and reference galaxies in identical
distance intervals shown in Figure~8.}
\halign { # \hfil & # \hfil  & \hfil # \hfil\cr
\noalign {\smallskip}
\noalign{\hrule}
\noalign{\smallskip}
1st NN distribution~~~& 2nd NN distribution~~~ & $\beta$ \cr
\noalign{\smallskip}
\noalign{\hrule}
\noalign {\medskip}
 BCG4--RM4  &  RM4--RM4   & 0.157 \cr
 BCG4--RM4  &  RA4--RM4   & 0.677 \cr
 BCG4--RM4  &  RD4--RM4   & 0.664 \cr
 RM4--RM4   &  RA4--RM4   & 0.015 \cr
 RM4--RM4   &  RD4--RM4   & 0.000 \cr
 RA4--RM4   &  RD4--RM4   & 0.434 \cr
\noalign {\medskip}
 BCG6--RM6  &  RM6--RM6   & 0.015 \cr
 BCG6--RM6  &  RA6--RM6   & 0.063 \cr
 BCG6--RM6  &  RD6--RM6   & 0.106 \cr
 RM6--RM6   &  RA6--RM6   & 0.024 \cr
 RM6--RM6   &  RD6--RM6   & 0.001 \cr
 RA6--RM6   &  RD6--RM6   & 0.594 \cr
\noalign {\medskip}
 BCG8--RM8  &  RM8--RM8   & 0.109 \cr
 BCG8--RM8  &  RA8--RM8   & 0.090 \cr
 BCG8--RM8  &  RD8--RM8   & 0.104 \cr
 RM8--RM8   &  RA8--RM8   & 0.279 \cr
 RM8--RM8   &  RD8--RM8   & 0.069 \cr
 RA8--RM8   &  RD8--RM8   & 0.925 \cr
\noalign {\medskip}
 BCG10--RM10  &  RM10--RM10   & 0.014 \cr
 BCG10--RM10  &  RA10--RM10   & 0.027 \cr
 BCG10--RM10  &  RD10--RM10   & 0.029 \cr
 RM10--RM10   &  RA10--RM10   & 0.814 \cr
 RM10--RM10   &  RD10--RM10   & 0.693 \cr
 RA10--RM10   &  RD10--RM10   & 0.999 \cr
\noalign{\medskip}
\noalign{\hrule} } 
\endtab

Random samples can be used to substitute either test or reference
samples. We have made several tests by changing actual  particles with
random ones, results for  these tests are presented in Figures~1 and 5.
Figure~5 shows that median NN distances of randomly located BCGs exceed
median NN distances of actual BCGs over 3 times. Our tests show that
the difference between actual and random samples is the greater the
more are objects of the particular sample concentrated to filaments or
other compact systems (cf. Einasto \etal 1991). Real galaxies are
strongly concentrated to filaments, their distribution is very
different from the distribution of randomly located points. In this
paper we are interested in {\it actual} median NN distances between
various types of galaxies, thus we use for reference only samples of
real objects.

\titleb {NN distance distribution for normal galaxies}
Next, we compare the NN distribution for BCGs with those of test samples
of normal galaxies. As reference samples absolute magnitude limited samples 
are used from now on. Since the completeness of the samples varies with
distance, we shall perform respective comparison separately for samples
taken at different distance intervals. We present differential and
cumulative distributions of NN distances as well.

The results of our analysis are presented in Figures~7 and 8. In
Figure~7 we show the differential and in Figure~8 the cumulative
distribution of NN distances. Calculations have been made for a 
number of subsamples taken at different distance intervals from 
$0 - 40$ \Mpc\ to $0 - 100$ \Mpc\ ($n = 4, 6, 8$ and $10$ for samples
BCGn, RMn, RAn and RDn, see Table~3). In all cases the NN distance of 
galaxies of the test sample from galaxies of the absolute magnitude 
limited reference sample RM was calculated. To estimate the similarity 
of distributions we performed KS tests for all distribution pairs. 
Results of the KS test are given in Table~4.

The KS test shows that at the distance interval $0 - 40$ \Mpc\ the
probability that the BCG--RM and RD--RM sample pairs have been drawn from
the same parent distribution is about 66\%. At larger distance intervals
this probability becomes smaller being 11\% at the distance interval $0 -
60$ \Mpc\ and much smaller at still larger distance intervals. 

The frequency of NN distances in the case of BCG test samples at small
argument values is smaller than the respective values of NN
distributions for normal galaxy samples. This can be seen in 
Figures 7 and 8 where the BCG -- RM curves 
lie beneath the others for small $D_{nn}$.
This result indicates that at small distances BCGs have less near
neighbours than normal galaxies, and that BCGs avoid dense regions.
This result was also found by Vilchez (1995) who studied
star--forming dwarf galaxies in different environments.

With the increase of the sample depth (panel a) to d)~) the NN
distances also increase for all samples.  The reasons of this effect
are, again, the decrease of the spatial galaxy density and selection
effects.

No evidence of an excess of distant neighbours of BCGs is visible 
in Figures~7 and 8 similar to the one suggested by PULTG. 

\titlea {Conclusions}
We have investigated the distribution of normal galaxies in 
voids and compared the distribution of blue compact dwarf galaxies 
with the distribution of normal galaxies. 
Our main results are as follows.

\item{$\bullet$}
The relative distribution of faint galaxies in voids defined by
bright galaxies of different absolute magnitude is virtually identical.
Voids differ not only by magnitude of surrounding galaxies,
they cover different volume in depth and have different size.
We call this similarity as void hierarchy.

\item{$\bullet$}
Faint galaxies in voids are concentrated towards void boundaries. 
A certain fraction of them are located in central
regions of voids where they are not distributed
randomly but form systems not containing any bright galaxy.

\item{$\bullet$}
The spatial distribution of BCGs is rather close to the distribution of
normal dwarf galaxies. The median distance of the nearest neighbours of 
BCGs from normal galaxies is only about 0.7 \Mpc, similar to the distance 
between members of groups and filaments of galaxies. 

\item{$\bullet$}
The number of close neighbours of BCGs is smaller than the equivalent 
number for other dwarf galaxies. This could be an indication
that BCGs are outlying members of small systems of galaxies.
Nearest neighbour distances of BCGs in respect to bright galaxies
have much larger values.

\item{$\bullet$}
The presence of very isolated distant BGCs found in several 
studies is a selection effect due to observational difficulties 
to detect normal dwarf galaxies at large distances.

\acknow{ UL acknowledges financial support by Verbundforschung
Astronomie/Astrophysik through BMWF/DARA grant FKZ~50~0R~90045.
JE, ME and VL thank Astrophysikalisches Institut Potsdam,
and JE and ME  G\"ottin\-gen Universit\"ats Sternwarte
for hospitality where a considerable part of this work was done. The
study was supported by Estonian Science Foundation grant 182 and
International Science Foundation grant LDC 000. We thank the 
anonymous referee whose suggestions helped to improve the 
presentation of our results.  }

\begref{References}
\ref Balzano, V.A. \& Weedman, D.W. 1982, \apj{255}, L1 
\ref Einasto, J., J\~oeveer, M., Saar, E., 1980, MNRAS, 193, 353
\ref Einasto, J., Klypin, A. A., Saar, E., 1984, MNRAS, 206, 529
\ref Einasto, J., Einasto, M., Gramann, M., Saar, E., 1991, MNRAS, 248, 593 
\ref Freudling, W., 1995, A\&AS, 112, 429
\ref Freudling, W., Haynes, M.P., Giovanelli, R., 1988, AJ, 96, 1791
\ref Freudling, W., Haynes, M.P., Giovanelli, R.,1992,
 ApJS, 79 , 157
\ref Freudling, W., Martel, H., and Haynes, M.P., 1991, ApJ, 377, 344
\ref Hopp, U., 1994, {\it Proceedings of the ESO OHP workshop
``Dwarf Galaxies''}, eds G. Meylin and P. Paturel, 37
\ref Huchra, J.P., 1994, ZCAT (compilation of galaxy redshifts)
\ref Kirshner, R. P., Oemler, A., Schechter, P. L. \& Shectman,
S. A., 1981, \apj{248}, L57
\ref Lindner, U., Einasto, J., Einasto, M., Freudling, W.,
Fricke, K.J., Tago, E., 1995, \aa{301}, 329 (paper~I)
\ref   Lipovetsky, V.A. et al. 1995, (in preparation).
\ref Moody, J.W., Gregory, S.A., Kirshner, R.P., \& Mac Alpine,
G.M., 1987, \apj{314}, L33
\ref Peimbert, M. \& Torres-Peimbert, S., 1992, \aa{253}, 349
\ref Pustil'nik, S.A., Ugryomov A.V. and Lipovetsky, V.A., 1994, 
Astronomical and Astrophysical Transactions,  5, 75
\ref Pustil'nik, S.A., Ugryomov A.V., Lipovetsky, V.A.,
Thuan, T.X., and Guseva, N.G., 1995, ApJ, 443, 499 (PULTG)
%\ref   Rosenberg, J. L., Salzer, J. J., Moody J.W., 1994, AJ, 108, 1557
%\ref   Salzer, J. J., 1989, ApJ, 347, 152
\ref Szomoru, A., van Gorkom, J. H., Gregg, M., de Jong, R. S., 1993,
AJ, 105, 464
\ref Szomoru, A.,  1994, PhD Thesis, Groningen University
\ref Vennik, J., 1984, Tartu Astroph. Obs. Teated, 73, 3
\ref Vennik, J., 1986, Astr. Nachr., 307, 157
\ref Vilchez, J.M., 1995, AJ, 110, 1090
\ref Weinberg, D. H., Szomoru, A., Guhathakurta, P., van Gorkom, J. H.,
1991, \apj{372}, L13
\ref Zwicky, F., Wield, P., Herzog, E., Karpowicz M. and Kowal,
C.T., 1961-68.  Catalogue of Galaxies and of Clusters of
Galaxies, 6 volumes. Pasadena, California Inst. Tech.
\endref
\bye